\documentstyle[aps,epsf]{revtex}

\begin{document}

\draft

\title{Neutron Stars and Massive Quark Matter}

\author{A.\ Drago and U.\ Tambini}

\address{Dipartimento di Fisica, Universit\'{a} di Ferrara, and INFN,
Sezione di Ferrara, Via Paradiso 12, I-44100 Ferrara, Italy}

\author{M.\ Hjorth-Jensen}

\address{ECT*, European Centre for Theoretical
Studies in Nuclear Physics and Related Areas, Trento, Italy}

\maketitle

\begin{abstract}
Using the Color-Dielectric model to describe
quark confinement, including strange quarks and
accounting for
beta--equilibrium, we get masses for a static neutron star in the range
$1.3\leq M/M_{\odot}\leq 1.54 $ for a radius $R\approx 9$ km.
Rapid cooling through the direct URCA mechanism is obtained. The implications
for the composition of neutron stars is discussed.
\end{abstract}


The equation of state (EOS) for dense matter is central to calculations of
neutron--star properties, such as the mass range, the mass--radius
relationship, the crust thickness
and the cooling rate,
see
e.g.\  Refs.\ \cite{prl95,lrp93,wg93}. The same EOS is also crucial
in calculating the energy released in a supernova explosion.

The typical density range of a neutron star
stretches from central densities of the order of 5 to 10
times the nuclear matter saturation density $n_0=0.17$
fm$^{-3}$ to zero at the edge of the star. Clearly,
the relevant degrees of freedom will not be the same in the crust,
where the density is much smaller than $n_0$, and in the center
of the star where the density is so high that models based
solely on interacting
nucleons are questionable.
Data from
neutron stars indicate that the EOS
should probably be moderately stiff in order to
support maximum neutron star masses in a range
from approximately
$1.4 M_{\odot}$ to $1.9 M_{\odot}$ \cite{thorsett93}. In addition,
simulations of supernovae explosions seem to require an EOS which is
even softer. A combined analysis of the data coming from binary pulsar
systems and from neutron star formation scenarios can be found
in Ref. \cite{finn}, where it is shown that neutron star masses should fall
predominantly in the range $1.3\le M/M_{\odot}\le 1.6 $.

Several theoretical approaches to calculations of the EOS
have been considered.
The hypothesis that strange quark matter may be the absolute
ground state of the strong interaction \cite{wit84},
has been used by several authors in the investigation of the
possibility of interpreting pulsars as rotating strange stars
\cite{glen91,rscoo92}.
Other approaches introduce
exotic states of nuclear matter, such as
kaon
\cite{brown93a,thorson93,brown93b,muto93} or pion condensation
\cite{migdal90,prog93}.

The scope of this work is to
study properties of neutron stars like cooling rate,
total mass and radius employing a massive quark model, the
so--called Color--Dielectric model (CDM) \cite{pirner92,birse90,dfb95}.
A very important feature of matter, as described using the CDM model,
is that the deconfinement phase transition happens at a density
of the order of 2--3 times $n_0$, a density expected to be reached in
the interior of a neutron star.
Neutron stars, when studied using the CDM, consist of an exterior region
described, {\it e.g.}, by the equation of state of Friedman and
Pandharipande\cite{fp81}, and an interior region of quark matter.
Moreover, the effective quark mass in the CDM is always larger than
a value of the order of 100 MeV, hence chiral symmetry is broken and
the Goldstone bosons are relevant degrees of freedom. This is to be
contrasted with models like the MIT bag,
where quarks have masses of a few MeV.
We therefore expect the CDM to be relevant
for computing the cooling rate of neutron stars {\it via}
the URCA
mechanism, as suggested by Iwamoto \cite{iwa}.

The CDM is a confinement model which has been used with success
to study
properties of
single nucleons, such as structure functions\cite{barone}
and form factors\cite{ff}, or to describe the
interaction potential between two nucleons \cite{kurt}, or to investigate
quark matter\cite{dfb95,mitja}.
In particular,
it is possible, using the same set of parameters,
both to describe the single nucleon properties
and to obtain meaningful results
for the deconfinement phase transition \cite{dfb95}.
The aim here is therefore to see whether this model
gives reasonable predictions for neutron star observables.
%This may
%in turn allow us to say something about the composition
%of matter in the interior of a neutron star.

The Lagrangian\footnote{
Throughout this paper we set $G=c=\hbar=1$, where $G$ is the gravitational
constant.}
of the model is given by:
\begin{eqnarray}
     {\cal L} &=& i\bar \psi \gamma^{\mu}\partial_{\mu} \psi \nonumber\\
     &+&\!\!\!\sum_{f=u,d} {g_f\over f_\pi \chi} \, \bar \psi_f\left(\sigma
     +i\gamma_5\vec\tau\cdot\vec\pi\right) \psi_f
     +{g_s \over \chi} \, \bar \psi_s \psi_s       \nonumber
     \\
      &+&{1\over 2}{\left(\partial_\mu\chi\right)}^2
      -V \left(\chi\right) \\
     \label{eq:in1}
     &+&{1\over 2}{\left(\partial_\mu\sigma \right)}^2
     +{1\over 2}{\left(\partial_\mu\vec\pi\right)}^2
     -U\left(\sigma ,\vec\pi\right)   \nonumber\, ,
\end{eqnarray}
where $U(\sigma ,\vec\pi)$ is the ``mexican-hat'' potential, as in
Ref.\ \cite{NF93}.
The lagrangian ${\cal L}$
describes a system of interacting $u$, $d$ and $s$ quarks, pions, sigmas and
a scalar--isoscalar chiral singlet field $\chi$
whose potential $U(\chi)$ is given by
\begin{equation}
    V(\chi)={1\over 2}{\cal M}^2\chi^2.
    \label{eq:in2}
\end{equation}

The coupling constants are given by $g_{u,d}=g (f_{\pi}\pm \xi_3)$
and $g_s= g(2 f_K -f_{\pi})$, where $f_{\pi}=93$ MeV and $f_{K}=113$ MeV
are the pion and the kaon
decay constants, respectively, and $\xi_3=f_{K^\pm}-f_{K^0}=-0.75$ MeV.
These coupling constants depend
only on a single parameter $g$.
This parameter is
fixed, together with the mass ${\cal M}$ of the $\chi$ field
so as to reproduce
experimental nucleon mass and radius.
The $SU(3)_f$ version of the model has been
introduced by Birse and McGovern\cite{su3,su3col}.

When considering a single hadron,
confinement is obtained {\it via} the effective quark masses
$m_{u,d}=g_{u,d} \bar\sigma/(\bar\chi f_\pi)$ and
$m_s=g_s / \bar\chi$,
which diverge outside the nucleon.
Indeed, the classical fields
$\bar \chi$ and $\bar\sigma$ are solutions of the Euler--Lagrange equations
and
$\bar\chi$ goes asymptotically to zero at large distances.

The model parameters $g$ and ${\cal M}$ are fixed to
reproduce single nucleon properties.
In order to describe the single nucleon state, a double projection on linear
and angular momentum eigenstates has to be performed, see Ref.\cite{NF93}.
We will use the parameters $g=0.023$ GeV and
${\cal M}=1.7$ GeV, giving a nucleon isoscalar radius of 0.80 fm
(exp.val.=0.79 fm) and an average
delta--nucleon mass of 1.129 GeV (exp.val.=1.085 GeV). A similar set
of parameters has been used to compute structure functions \cite{barone}
and form factors \cite{ff}.

The quark matter (QM) phase is characterized by a constant value of the
scalar fields and by using plane waves to describe the quarks.
The total energy of QM in the mean field approximation reads
\begin{eqnarray}
      E_{QM}&=&6 V \!\!\sum_{f=u,d,s}\int\!\!\!
      {d{\bf k} \over (2\pi)^3}
      \sqrt{{\bf k}^2 + m_f^2} \,
      \theta (k_F^f-k) \nonumber \\
      &+& V U(\bar \chi) + V W(\bar \sigma,\vec\pi=0),
      \label{eq:in4}
\end{eqnarray}
where $k_{F}^f$ is the Fermi momentum of quarks with flavour $f$.

The high--density matter in the interior of neutron stars is described by
requiring the system to be globally neutral
\begin{equation}
    \label{eq:neut}
    (2/3)n_u -(1/3)n_d - (1/3)n_s - n_e = 0,
\end{equation}
where $n_{u,d,s,e}$
are the densities of the $u$, $d$ and $s$ quarks and of the
electrons, respectively.
Morover, the system must be in $\beta$--equilibrium, i.e.\
the chemical potentials have to satisfy the following equations:
\begin{eqnarray}
      \label{eq:ud}
      \mu_d&=&\mu_u+\mu_e,\\
      \label{eq:us}
      \mu_s&=&\mu_u+\mu_e .
\end{eqnarray}
Eqs.\ (\ref{eq:neut})--(\ref{eq:us}) have to be solved
self--consistently together with
field equations, at a fixed baryon density $n=n_u+n_d+n_s$.

\begin{figure}
       \setlength{\unitlength}{1mm}
       \begin{picture}(140,150)
       \put(25,10){\epsfxsize=12cm \epsfbox{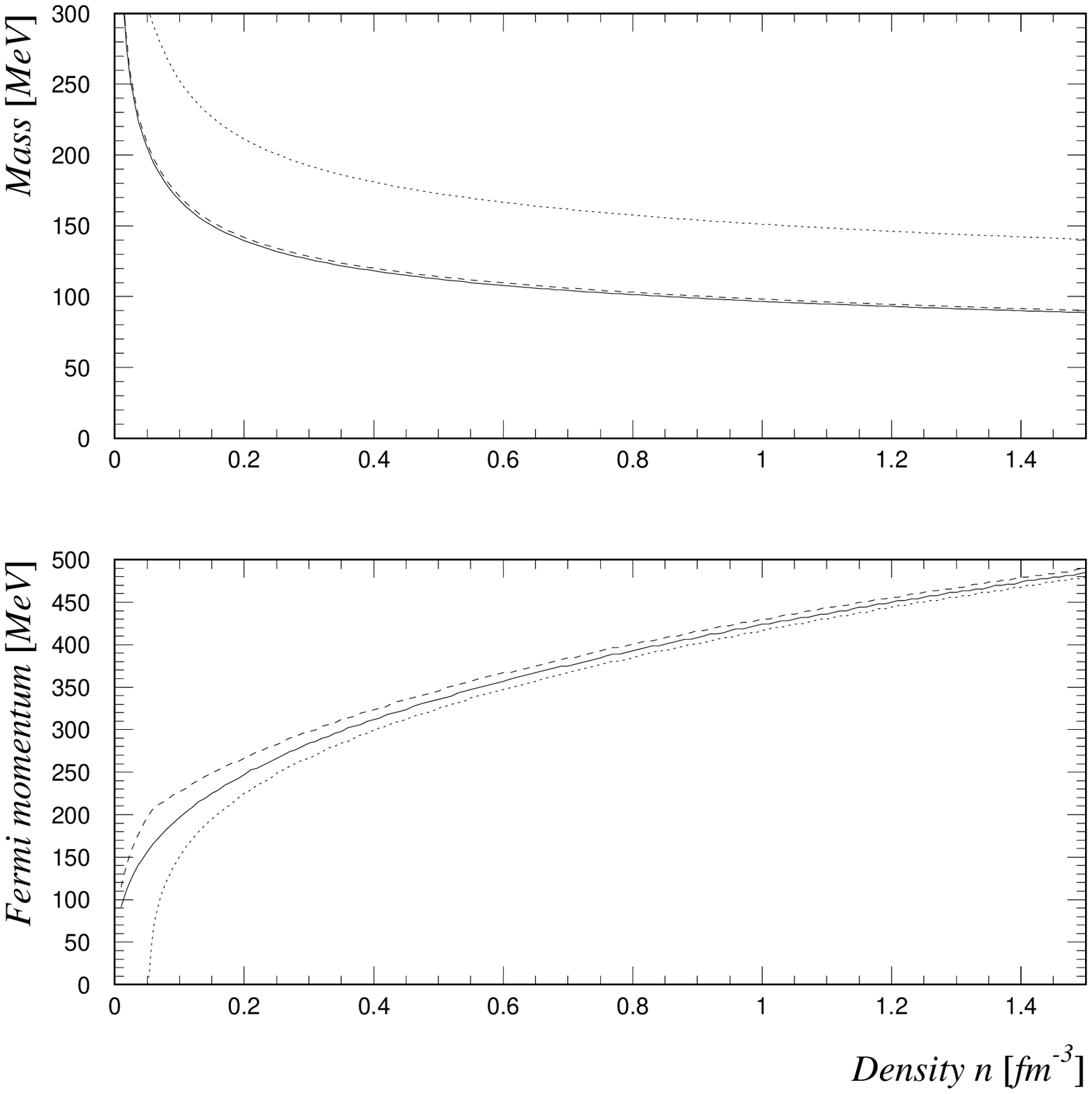}}
       \end{picture}
       \caption{The upper part shows quark masses as
                function of density $n$. In the lower figure we give
                the quark composition as function of density $n$. In both
                figures, solid lines are the results for $u$ quarks, dashed
                lines correspond to $d$ quarks and dotted lines are the
                results for $s$ quarks.}
       \label{fig:fig1}
\end{figure}

In the upper part of Fig.\ \ref{fig:fig1},
the resulting quark masses are shown as function of density.
As can be seen, the quark masses are large and of the order of 100 MeV
up to densities about 10 times $n_0$.
In the lower part of Fig.\ \ref{fig:fig1},
the Fermi momenta of the quarks are shown. In the relevant
region for neutron stars the momenta are of the order of 400 MeV.
Strange quarks appear at a density of $\simeq 1/3 n_0$.

\begin{figure}
       \setlength{\unitlength}{1mm}
       \begin{picture}(140,150)
       \put(25,10){\epsfxsize=12cm \epsfbox{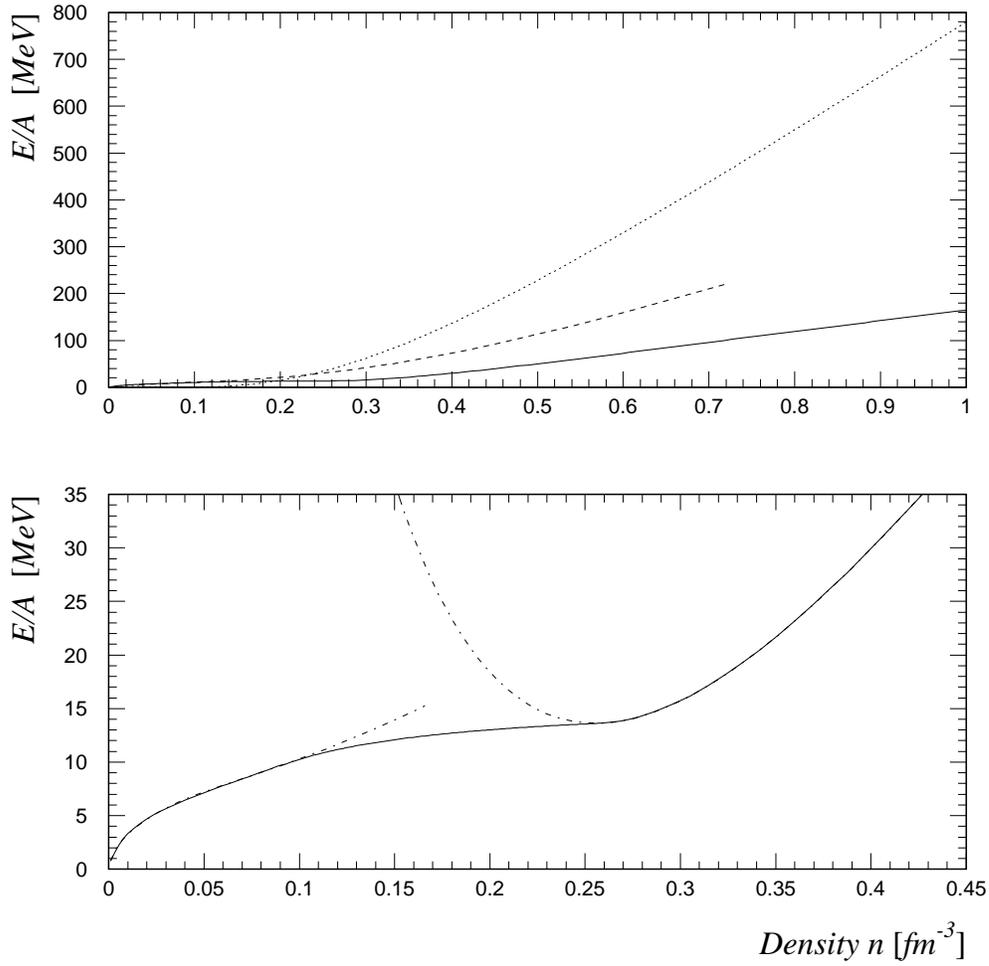}}
       \end{picture}
       \caption{The upper part gives the energy per baryon $E/A$
                as function of the density $n$ for the three
                equations of state discussed in the text.
                The solid line corresponds
                to the result obtained with the CDM,
                the dotted
                line represents the result for the Walecka model,
                while the dashed line is the result from the DBHF
                calculations of Ref.\ [28]. In the lower figure,
                we show
                the Maxwell construction discussed in the text.
                The solid line
                is the final EOS after the Maxwell construction,
                the dash-dotted lines to the left
                (low densities) is the EOS of Ref.\ [18], while the dash-dotted
                line to the right is the CDM EOS.}
                \label{fig:fig2}
\end{figure}

The energy of the QM, as described by the CDM, is larger than the energy
of nuclear matter for densities smaller than 2--3 times $n_0$.
%Since the EOS derived from the CDM is valid at densities above the
%saturation density of nuclear
%matter, we need to link this EOS with equations of state which
%are more appropriate for densities below 2--3 times $n_0$.
For lower densities
we take the results from Friedman and Pandharipande \cite{fp81},
listed in Table 5 of Ref.\ \cite{prl95}.
We obtain an EOS valid at all densities by joining
the EOS of \cite{fp81} and the CDM EOS through
a Maxwell construction. The latter starts at a density $\simeq
0.1 {\rm fm}^{-3}$
and terminates at a density $\simeq 0.26 {\rm fm}^{-3}$.
The energy per baryon $E/A$ in beta--stable nuclear matter
is displayed in the upper part of
Fig.\ \ref{fig:fig2} as function
of density $n$. In the lower part of Fig\ \ref{fig:fig2}
we show the above Maxwell construction. For the sake of comparison we also
include the results from pure neutron matter obtained
with the so--called Walecka model \cite{sw86} and the energy per
particle obtained from a relativistic Dirac--Brueckner--Hartree--Fock
(DBHF) calculation in neutron matter \cite{behoo94}, using
modern meson-exchange potential models \cite{mac89}.
As can be seen from Fig.\ \ref{fig:fig2}, our
CDM EOS is much softer than those obtained
with either the Walecka model or the DBHF calculation.

{}From the general theory of relativity,
the structure of a static neutron star is determined through the
Tolman--Oppenheimer--Volkov equation, i.e.\
\begin{equation}
   \frac{dP}{dr}=
    - \frac{ \left\{\rho (r)+P(r) \right\}
   \left\{M(r)+4\pi r^3 P(r)\right\}}{r^2- 2r M(r)},
   \label{eq:tov}
\end{equation}
where $P(r)$ is the pressure and $M(r)$ is
the gravitational mass inside a radius $r$.
The pressure is related to the energy per particle
${\cal E} =E/A$ by the
relation
$P(n)=n^2 \left(\partial {\cal E}/\partial n\right)$.

\begin{figure}
       \setlength{\unitlength}{1mm}
       \begin{picture}(140,150)
       \put(25,10){\epsfxsize=12cm \epsfbox{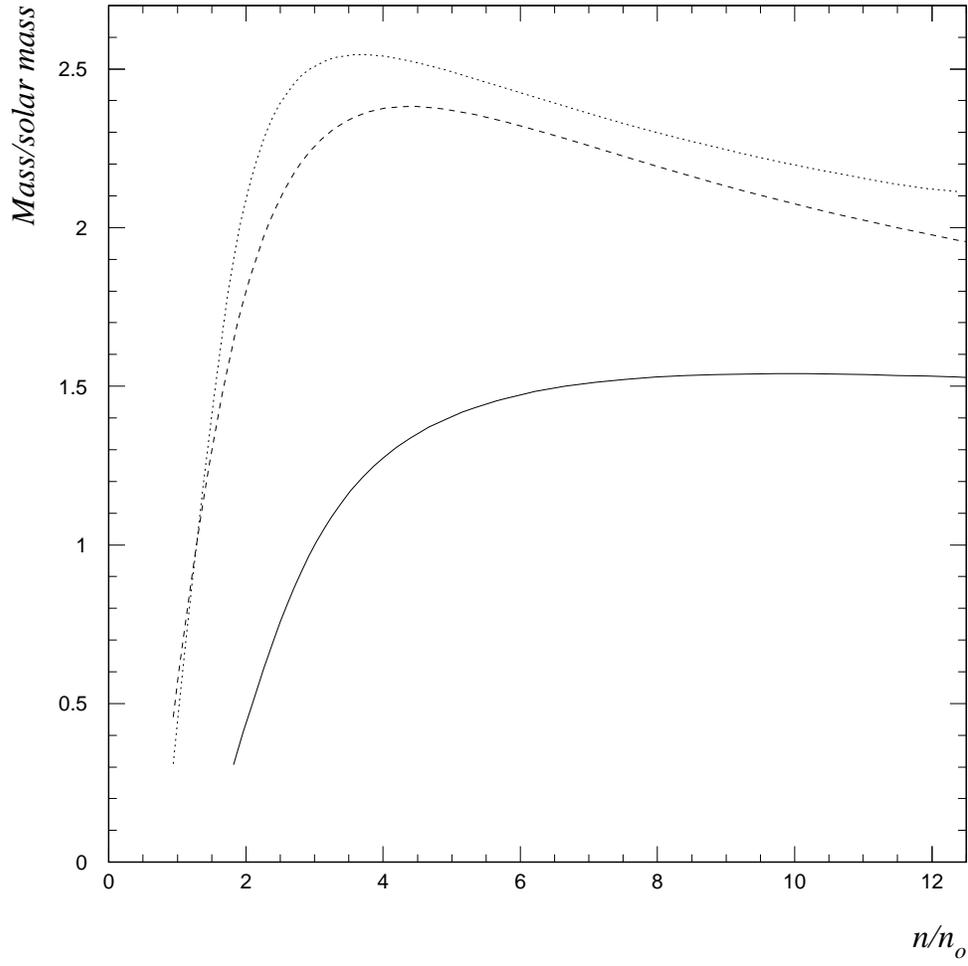}}
       \end{picture}
       \caption{$M/M_{\odot}$ as function of $n/n_0$, where
                $n_0$ is the nuclear matter saturation density $n_0=0.17$
                fm$^{-3}$. The solid line corresponds
                to results obtained with the CDM,
                the dotted
                line represents results for the Walecka model,
                while the dashed line are results from the DBHF
                calculations of Ref.\ [28]. }
                \label{fig:fig3}
\end{figure}

\begin{figure}
       \setlength{\unitlength}{1mm}
       \begin{picture}(140,150)
       \put(25,10){\epsfxsize=12cm \epsfbox{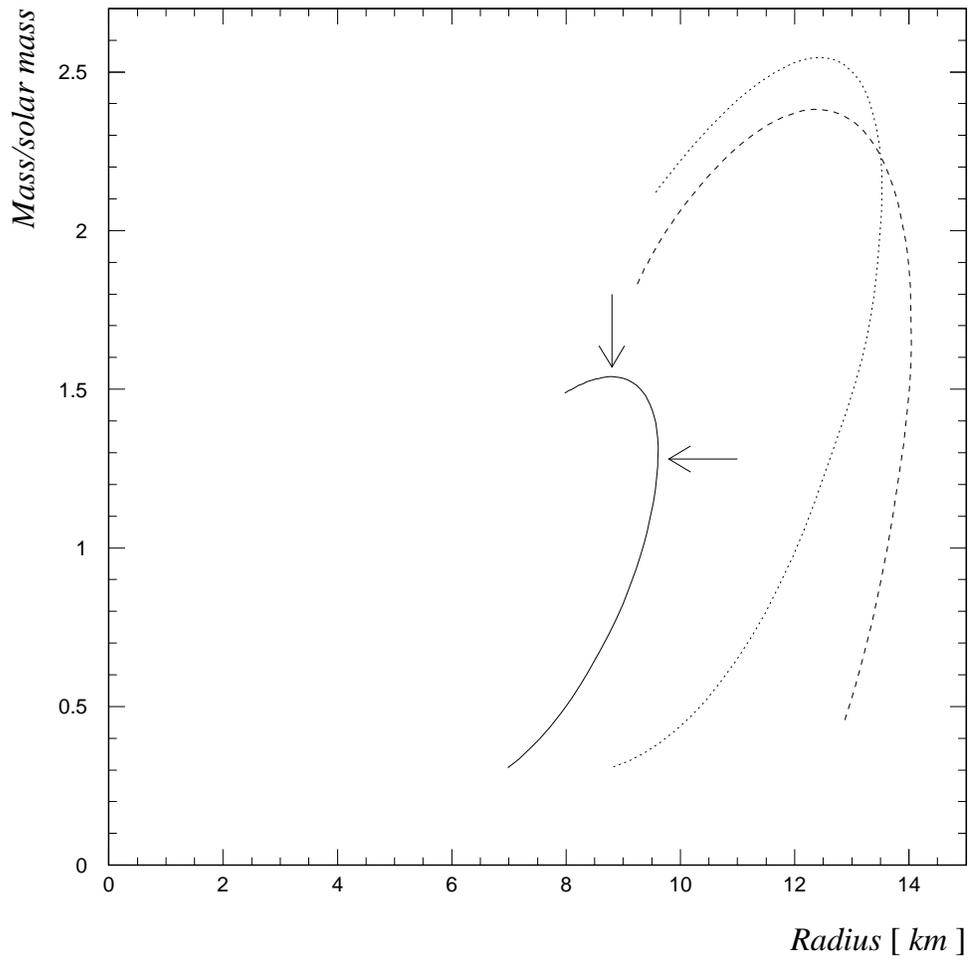}}
       \end{picture}
       \caption{Total mass
                $M/M_{\odot}$ as function of $R$. Notations are the
                same as in Fig.~3.
                The arrows indicate the maximum and minimum mass obtained with
                the CDM EOS.}
                \label{fig:fig4}
\end{figure}

The resulting mass
is shown in Fig.\ \ref{fig:fig3}, while
the total mass as function of radius is displayed in Fig.\ \ref{fig:fig4}.
With the CDM we obtain a maximum mass
$M_{\rm{max}}\approx 1.54 M_{\odot}$
and a radius of 9 km at a central density
corresponding to approximately 9 times nuclear matter
saturation density, in good agreement with the experimental values
for the mass \cite{thorsett93,finn} and the estimate for the radius
\cite{mo88}.
Concerning the minimum mass, we obtain a value equal to $1.3 M_{\odot}$.
It is also possible to compute the minimum period of rotation of the star,
given by $P_m\sim (0.8$ms$) R_{10}^{3/2}(M_{\odot}/M)^{1/2}$,
where $R_{10}$ is the radius in units of 10 km\cite{haensel}. In our
model we get $P_m=0.55$ ms.

Finally, we study the neutrino and antineutrino emission,
which can be responsible for rapid cooling of neutron stars (URCA
mechanism).
In the interior of the star, where, in our model, strange quark matter
is present, the relevant reactions are:
\begin{eqnarray}
     d&\rightarrow &u + e^- + \bar\nu_e, \\
     e^- + u &\rightarrow & d + \nu_e, \\
     s&\rightarrow &u + e^- + \bar\nu_e, \\
     e^- + u &\rightarrow & s + \nu_e .
\end{eqnarray}
To conserve momentum in the reactions, the following inequalities have to
be satisfied:
\begin{eqnarray}
    \label{kud}
    \vert k_F^u - k_F^e \vert &\leq & k_F^d \leq  k_F^u + k_F^e, \\
    \vert k_F^u - k_F^e \vert &\leq & k_F^s \leq  k_F^u + k_F^e.
    \label{kus}
\end{eqnarray}

For densities larger than 0.53 fm$^{-3}$, conditions (\ref{kud}) are satisfied
and the direct URCA mechanism involving only $u$ and $d$ quarks can start.
Conditions (\ref{kus}) are satisfied
only for densities larger than 1.4 fm$^{-3}$. The URCA mechanism
involving strange quarks is
suppressed by a factor $\sin ^2\theta_c\simeq 0.05$ with respect to the
previous process,
where $\theta_c$ is the Cabibbo angle.

We compute the neutrino and antineutrino luminosity, using the
ultrarelativistic expansion for the chemical potentials\cite{iwa}. This
approximation is reasonable because the ratio $m_q/k_F$ is approximately equal
to 0.25.
One gets for the total luminosity \cite{iwa}
\begin{equation}
   \epsilon={457\over 1680} G_F^2 \cos^2\theta_c
    m_d^2 f k_F^u (k_B T)^6,
\end{equation}
where $f=1-(m_u/m_d)^2 (k_F^d/k_F^u) - (m_e/m_d)^2 (k_F^d/k_F^e)$.
To obtain the characteristic cooling time we equate the energy loss per
unit volume to the rate of change of thermal energy per unit volume
$\tau=c_V T/\epsilon$. Here
the heat capacity of the QM is $c_V=\sum_{f=u,d,s} m_f k_F^f k_B^2 T$.
We get $\tau=C\,\, 1$day$/T_9^4$, where $T_9$ is the temperature measured in
units of $10^9 \,^{\rm o} {\rm K}$
and $C$ is a constant ranging from 0.5 to $\sim$ 10
going from heavy to light neutron stars.
To compare with experiment, one should also account for
possible reheating mechanisms. There are however indications\cite{umeda}
that temperatures of young ($\sim 10^4$ years old) neutron stars lie
below that obtained through the indirect URCA processes. This suggests
that cooling through exotic mechanisms, as accounted for in our
model, may be a meaningful explanation.
Moreover, a recent reappraisal of neutrino--pair bremsstrahlung
by Pethick and Thorsson \cite{pt94} indicates that this process
is much less important for the thermal evolution of neutrons stars than
suggested by
earlier calculations.
In addition, since the results of Ref.\ \cite{prl95,lrp93} indicate that
the crust of the neutron star is considerably smaller than previously
extimated, and the results of Ref.\ \cite{deehlo95} show that proton
superconductivity in the stellar interior is substantial, thereby inhibiting
the traditional URCA mechanisms, there is a possibility that exotic states of
matter like QM provide a viable explanation for various properties of neutron
stars.

In summary, here we show that the CDM,
which has been successful in describing various
properties of hadrons, also gives reasonable results for neutron star
observables like the mass and radius. Further, our results for the cooling
of the star, combined with the results of
Refs.\ \cite{prl95,lrp93,pt94,deehlo95},
may suggest that the core of the neutron star can be described by
a QM phase.
Lastly, supernovae simulations require an EOS which is soft
\cite{bethe90}, like ours.

This work has been supported by the Istituto Nazionale di Fisica
Nucleare (INFN) , Italy, the Istituto Trentino di Cultura, Trento, Italy,
and the Research Council of Norway. We are also indebted to Profs.\
Luca Caneschi, D\ .Mukhopadyay and Erlend \O stgaard for critically reviewing
the manuscript.

%           References

\clearpage

\end{document}